\documentclass[
reprint,
superscriptaddress,
nofootinbib,
amsmath,
amssymb,
aps,
prl,
twocolumn
]{revtex4-2}
\usepackage{orcidlink}
\usepackage{graphicx}
\usepackage{dcolumn}
\usepackage{bm}
\usepackage{hyperref}
\usepackage{xcolor}
\usepackage{subfigure}
\usepackage{subcaption}
\usepackage{orcidlink}
\usepackage{txfonts}
\usepackage{booktabs}
\usepackage{mathtools}
\usepackage{ragged2e}
\usepackage{caption}
\usepackage{todonotes}
\usepackage{xr}
\usepackage[titletoc, title]{appendix}
\usepackage[percent]{overpic}
\usepackage{adjustbox}
\usepackage{epstopdf}
\usepackage{titlesec}
\usepackage{listings}
\usepackage{verbatim}
\usepackage{subcaption}
\usepackage{ulem}

\begin{document}

\title{Search for Light-Mass Fractionally Charged Particles in Space with DAMPE Experiment}

\author{F.~Alemanno,\orcidlink{0000-0003-2021-9205}}
\affiliation{Dipartimento di Matematica e Fisica E. De Giorgi, Universit\`a del Salento, I-73100, Lecce, Italy}
\affiliation{Istituto Nazionale di Fisica Nucleare (INFN) - Sezione di Lecce, I-73100, Lecce, Italy}

\author{Q.~An}
\altaffiliation[]{Deceased}
\affiliation{State Key Laboratory of Particle Detection and Electronics, University of Science and Technology of China, Hefei 230026, China} 
\affiliation{Department of Modern Physics, University of Science and Technology of China, Hefei 230026, China}
\author{P.~Azzarello}
\affiliation{Department of Nuclear and Particle Physics, University of Geneva, CH-1211, Switzerland}

\author{F.~C.~T.~Barbato,\orcidlink{0000-0003-0751-6731}}
\affiliation{Gran Sasso Science Institute (GSSI), Via Iacobucci 2, I-67100 L'Aquila, Italy}
\affiliation{Istituto Nazionale di Fisica Nucleare (INFN) - Laboratori Nazionali del Gran Sasso, I-67100 Assergi, L'Aquila, Italy}

\author{P.~Bernardini,\orcidlink{0000-0002-6530-3227}}
\affiliation{Dipartimento di Matematica e Fisica E. De Giorgi, Universit\`a del Salento, I-73100, Lecce, Italy}
\affiliation{Istituto Nazionale di Fisica Nucleare (INFN) - Sezione di Lecce, I-73100, Lecce, Italy}

\author{X.~J.~Bi}
\affiliation{University of Chinese Academy of Sciences, Beijing 100049, China}
\affiliation{Particle Astrophysics Division, Institute of High Energy Physics, Chinese Academy of Sciences, Beijing 100049, China}

\author{H.~V.~Boutin,\orcidlink{0009-0004-6010-9486}}
\affiliation{Department of Nuclear and Particle Physics, University of Geneva, CH-1211, Switzerland}

\author{I.~Cagnoli,\orcidlink{0000-0001-8822-5914}}
\affiliation{Gran Sasso Science Institute (GSSI), Via Iacobucci 2, I-67100 L'Aquila, Italy}
\affiliation{Istituto Nazionale di Fisica Nucleare (INFN) - Laboratori Nazionali del Gran Sasso, I-67100 Assergi, L'Aquila, Italy}

\author{M.~S.~Cai,\orcidlink{0000-0002-9940-3146}}
\affiliation{Key Laboratory of Dark Matter and Space Astronomy, Purple Mountain Observatory, Chinese Academy of Sciences, Nanjing 210023, China}
\affiliation{School of Astronomy and Space Science, University of Science and Technology of China, Hefei 230026, China}

\author{E.~Casilli,\orcidlink{0009-0003-6044-3428}}
\affiliation{Gran Sasso Science Institute (GSSI), Via Iacobucci 2, I-67100 L'Aquila, Italy}
\affiliation{Istituto Nazionale di Fisica Nucleare (INFN) - Laboratori Nazionali del Gran Sasso, I-67100 Assergi, L'Aquila, Italy}

% ...
\collaboration{DAMPE Collaboration}
\altaffiliation{dampe@pmo.ac.cn}

\date{\today}

\begin{abstract}

Free Fractionally Charged Particles (FCPs) are predicted by some theories beyond or extended to the standard model. FCPs have been widely searched for by underground and space-based experiments based on the assumption of heavy lepton-like particles. However, there is a paucity of research focusing on light-mass FCPs (LFCPs) in the sub-MeV mass range. In this work, we report the LFCPs in primary high energy cosmic rays, based on observational data from the Dark Matter Particle Explorer (DAMPE) satellite. This study utilized ten years on-orbit data of DAMPE to search for LFCPs with a charge of $\frac{2}{3}~e$. No LFCP candidate was observed. 
Upper flux limit of LFCPs with a mass of 0.511 MeV$/c^{2}$ and a charge of $\frac{2}{3}~e$ is determined to be $\rm 5.0 \times 10^{-11}\,cm^{-2}sr^{-1}s^{-1}$ at the $\rm 90\%$ confidence level.

\end{abstract}


\maketitle

$Introduction$ - In the 1960s, the quark model was proposed by Gell-Mann and Zweig~\cite{Gell-Mann:1964ewy, Zweig:1964jf}. This model introduced fractionally charged constituents (with a charge of $\pm \frac{1}{3}\,e$ or $\pm \frac{2}{3}\,e$) to explain the composition inside the hadron. However, the Quantum ChromoDynamics incorporates color confinement, which forbids the isolation of free quarks~\cite{gross1973ultraviolet, bander1981theories}. Within the framework of the Standard Model (SM) theory, all free fundamental particles have integer charges of 0 or $\rm \pm1$.
Nevertheless, various extensions to the SM theory, including grand unification~\cite{ParticleDataGroup:2022pth}, super-symmetry~\cite{Frampton:1982gc}, and certain astrophysical production mechanisms, allow for Fractionally Charged Particles (FCPs) that could be a kind of free particle with charges of quarks or far less than $e$ named as millicharge~\cite{Jones:1976xy, Dobroliubov:1989mr, Cecchini:1993ez, Glashow:2005jy}. Detection of such particles could also provide new candidates for dark matter \cite{Kouvaris:2013gya, Shiu:2013wxa, Arza:2025cou}.

Over the past decades, a variety of experiments have searched for the FCPs. Accelerator-based collider or fixed-target experiments have been conducted to study FCPs that extend or transcend the SM theory~\cite{Lyons:1986aq, Prinz:1998ua, ArgoNeuT:2019ckq, Ball:2020dnx, SENSEI:2023gie, CMS:2024eyx}. 
These experiments provide a controlled, reproducible environment with well-understood production mechanisms and backgrounds, yet they are constrained by finite energy reach, small production cross sections, and residual model dependence.
Alternatively, searching for FCPs in cosmic rays is also an important experimental direction, since the FCPs could be produced (a) at the early universe, (b) by the violent astrophysical activities in space, (c) or during the process of the extensive air shower~\cite{Perl:2004qc}. The variety of production conditions results in an undefined search range for the mass.

The assumption of heavy lepton-like FCPs with the mass beyond several MeV$/c^{2}$ is widely adopted by searches in cosmic rays, of which the ionization dominants the energy loss process and the bremsstrahlung effect is suppressed. Behaving like the Minimum Ionization Particles (MIPs), the heavy lepton-like FCPs could go through the interstellar medium, cross the atmosphere, or penetrate the crustal materials with high kinetic energy. These searches include the underground experiments\cite{Kamiokande-II:1990hos, Aglietta:1994iv, MACRO:2000bht, CDMS:2014ane, Majorana:2018gib, SuperCDMS:2020hcc, CUORE:2024rbd}, ground-based experiments~\cite{hsiao1982search,Gomez:1967zz,Mccusker:1969ep}, bulk-matter searches~\cite{smith1989searches, halyo2000search,Pospelov:2020ktu, Afek:2020lek}, balloon-borne \cite{Fuke:2008zza} and space-based facilities~\cite{Sbarra:2003ur, DAMPE:2022zzr}.

With respect to the assumption of the Light-mass FCPs (LFCPs) with smaller mass of sub-MeV, the bremsstrahlung effect would increase predominantly, 
as it induces the cascade processes, like electron-positron pair productions, when interacting with materials. 
This cause the LFCPs in cosmic rays to be readily absorbed during propagation, thereby hindering their penetration of the atmosphere or crustal materials, and making them difficult to detect with (under)ground-based facilities.
So, there are few studies on the LFCPs which could be generated by possible nearby celestial activities, while on-orbit experiments have great advantages in this field.

The DArk Matter Particle Explorer (DAMPE) \cite{DAMPE:2017cev, DAMPE:2019lxv}, also known as "Wukong", launched on 17 December 2015, is an on-orbit satellite-borne calorimetric-type detector aiming to indirectly search for dark matter by exploring the high-energy cosmic rays in space. DAMPE has been stably operating for more than ten years, accumulated nearly 20 billion high-energy events of cosmic rays. Based on the numerous scientific data, DAMPE has released important scientific contributions of cosmic-rays measurements~\cite{DAMPE:2017fbg,DAMPE:2019gys,Alemanno:2021gpb,DAMPE:2022zzr,DAMPE:2022jgy, DAMPE:2025opn}.

DAMPE consists of four sub-detectors. A Plastic Scintillator Detector (PSD) is made of two layers of 41 plastic scintillator units\cite{Yu:2017dpa, Ding:2018lfn}. A Silicon-Tungsten tracKer converter (STK) is composed with six modules of double-layer silicon strip detectors \cite{Azzarello:2016trx, DAMPE:2017yae,qiao2019charge}. A Bismuth Germanium Oxide (BGO) imaging calorimeter has fourteen layers of 22 crystal bars~\cite{Zhang:2016xkz, Wu:2018gyd} with the capability of three-dimensional imaging for electromagnetic and hadronic showers. And a NeUtron Detector (NUD)\cite{Huang:2020skz} consists of four boron-doped plastic scintillation tiles. The PSD and STK are both used to measure the charges of incident particles. The STK, in conjunction with the track provided by the BGO calorimeter, reconstructs the precise trajectories of particles. The BGO calorimeter is used primarily to detect the energy of incident particles. The topology of the shower profiles in the BGO calorimeter can effectively separate electromagnetic particles and hadrons. The BGO calorimeter also provides the trigger for DAMPE detector. The NUD is used for further electron/proton discrimination. 

The charge resolutions of PSD and STK are 0.06 $e$~\cite{Dong:2018qof} and 0.04 $e$~\cite{Vitillo:2017zbj}, respectively, for measuring the singly charged particles. The high-precision performance of the BGO calorimeter provides a reliable energy measurement of incident particle. DAMPE has the capability to identify the LFCPs and record its deposited energy. 
In this letter, a research of LFCPs with charge of $\frac{2}{3}\,e$ and masses in the range $0.2\ \mathrm{MeV}/c^2$ – $1\ \mathrm{MeV}/c^2$ was conducted, using data collected by DAMPE in space over a period of ten years. 

$Data\;Analysis$ - This work uses the data collected by DAMPE detector from January 1, 2016 to December 31, 2025. The time when the satellite flying over the South Atlantic Anomaly region~\cite{DAMPE:2019lxv} is excluded.
Periods of instrumental dead time and on-orbit calibration are also removed. 
Furthermore, the data from September 9 to September 13, 2017 are excluded because of a giant solar flare.
Thus, the total effective exposure time is about $\rm 2.4 \times 10^8\,s$, about 76\% of the total operational duration.

Based on the Geant4 toolkit~\cite{GEANT4:2002zbu}, a model of LFCPs is created within the DAMPE software considering the physical processes of ionization, bremsstrahlung, and multiple scattering. The charge value is set to be $\frac{2}{3}\,e$, since the ionized energy is proportional to the square of charge, and smaller charges of LFCPs result in lower $\rm dE/dx$ (the gradient of ionization energy loss), which leads to a poorer signal/noise ratio and less efficiency of track reconstruction of STK. In addition, the mass of LFCPs was set varying from 0.2 MeV$/c^{2}$ to 1 MeV$/c^{2}$. 
Further information of establishing the LFCPs model is introduced in the Supplemental Materials. 
The Monte Carlo (MC) simulations of electrons, protons are introduced to evaluate the background. The MC samples are generated from a hemispherical isotropic source obeying the power-law energy spectrum. The index of the energy spectrum is adjusted from -1 to -2.7 following the general cosmic-rays spectrum. 

The major LFCP samples analyzed in this work carry mass of 0.511 MeV$/c^{2}$. As the assumed mass of LFCPs decreases, the bremsstrahlung effect increases, causing poor charge resolution for both PSD and STK.

$(i)\;Preselection$ - A series of requirements are imposed to primarily extract pure electron-like particles. 
(a) Fiducial. Based on the energy barycenter in each layer of BGO calorimeter, a BGO track is reconstructed. 
An event with a BGO track must traverse the whole DAMPE detector without penetrating the edges of the first layer of PSD and the last layer of BGO calorimeter to ensure good containment of the shower. 
A cut on the maximum-layer to total energy fraction (\textless 35\%) is applied to ensure the rejection of events incident from the side of the DAMPE detector. 
(b) Hadrons separation. The three-dimensional topological structure of shower profiles in BGO is used to exclude the hadrons~\cite{DAMPE:2017fbg}.
(c) Initial energy selection. Due to the geomagnetic cutoff effect, the deposited energy of candidate events should be above 10 GeV.

$(ii)\;Trigger\;selection$ - The selected events are required to satisfy the High Energy Trigger (HET) pattern~\cite{Zhang:2019wkj} that is designed to select the high-energy showered events. Due to the bremsstrahlung effect, the LFCPs are possible to deposit large energy activating the HET. The HET demands the maximum energy deposition among 22 crystals larger than 10 times of proton MIPs energy (one proton MIP deposits about 23 MeV in one BGO crystal) in each of the first three BGO layers, and larger than 2 MIPs in the fourth layer. Although the LFCPs are assumed to be electron-like, the trigger efficiency of LFCPs is lower than that of electrons since the lower energy deposition in the early stage of the shower. The trigger efficiencies are about 77.9\%, 92.2\% and 90.17\% for MC LFCPs, MC electrons and on-orbit data, respectively.

$(iii)\;Track\;selection$ - A high-quality trajectory is selected from the STK among the trajectories reconstructed by the Kalman filter algorithm~\cite{DAMPE:2017yae} as STK track within a distance of 15 mm form the BGO track. This track is evaluated by the hits number in sub-layers and the $\rm \chi^2/ndof$ of the Kalman fit. Compared with electrons, LFCPs are more sensitive to the interference of secondary particles during STK track reconstruction, which results in a broader track $ \rm \chi^2$ distribution. Accompanying with the less $\rm dE/dx$ in silicon strips, LFCPs have a lower reconstruction efficiency which is about 88.4\%, and 97.8\% for MC electrons, along with 94.60\% for on-orbit data.

$(iv)\;Charge\;selction$ - Both of the PSD and STK detectors are used to reconstruct charges of LFCPs. A set of selections are utilized to enhance the accuracy and the reliability of the charges. (a) The PSD strip is required to be penetrated by STK track from top to bottom surfaces at each of the two PSD layers, which are used to reconstruct the PSD charges corrected with the path length, light attenuation, and quenching effects~\cite{Ding:2018lfn}. Signals from both ends of the strip are required to be consistent by constraining the ratio of the ends' charge values. 
The distribution of two-end ratios is fitted with a gaussian function and limited to be within the range of three times the width around the peak.
(b) To purify the signal at STK layers and eliminate the influence of secondary particles, the first four sub-layers silicon detectors are used. The deposited energy of each sub-layer along the track is corrected with the hit position~\cite{DAMPE:2021pwi} and path length. The strips number and deposited energy of a signal are strictly constrained. 
At least three effective signals from STK detector is required to decrease the charge fluctuation.
Details of the charge selections are described in the Supplemental Materials.

The PSD and STK detectors are utilized independently to reconstruct the charges of selected events in this work. The average of charge from the two layers of the PSD is adopted as the final PSD charge value. The average of the remaining effective signals of the STK is adopted as the STK charge. The charge smearing is applied to maintain the consistence between the MC electrons sample and on-orbit data. The MC LFCP sample is smeared with the same parameters of MC electrons.

The efficiencies of charge selection consist the contributions from the PSD and STK detectors. Separately, efficiencies of PSD and STK for MC LFCPs are 92.0\% and 42.3\%, for MC electrons 98.0\% and 37.8\%, and for on-orbit data 96.4\% and 36.9\%. The strict criteria of STK charge reconstruction results in the low efficiency. In total, efficiencies for MC LFCPs, MC electrons and on-orbit electron data are 38.9\%, 37.0\% and 35.5\%, respectively.

\begin{figure}[h]
	\centering
    \begin{overpic}[width=0.45\textwidth]{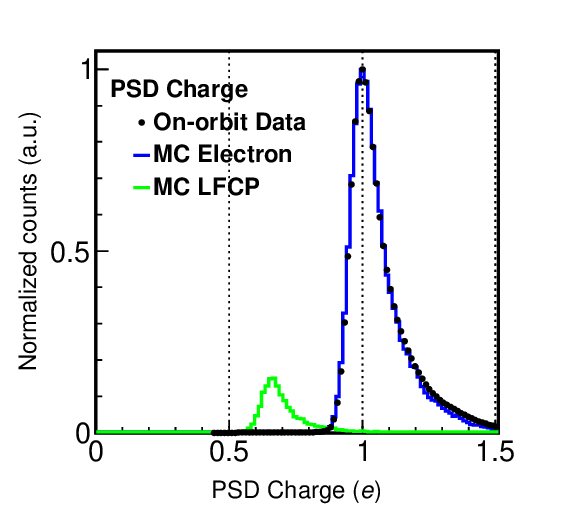}
        \put(2,80){\Large{(a)}}
    \end{overpic}
    \begin{overpic}[width=0.45\textwidth]{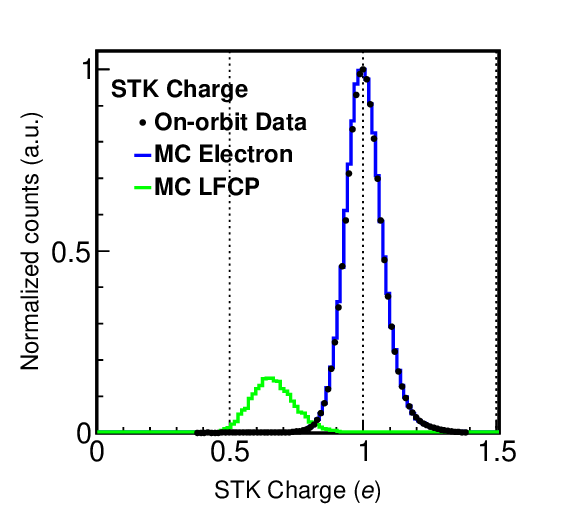}
        \put(2,80){\Large{(b)}}
    \end{overpic}
	\caption{The distributions of charges measured by the PSD (a) and STK (b). A Landau-Gaussian convoluted function was used to fit the on-orbit data (black points), MC electrons (blue line) , and MC LFCPs (green line) separately.  \label{fig:Fitting}}
\end{figure}

The PSD and STK charge spectra of on-orbit data, MC electrons and LFCPs with 0.511 MeV$/c^{2}$ mass are depicted in the Fig.~\ref{fig:Fitting}. MC electrons spectrum has a good agreement with on-orbit data, and the LFCPs charges are apparently separated from singly charge particles.

$(v)\;Effective\;acceptance$ - The effective acceptance resulted from the combined effects of all the mentioned criteria and is estimated with the MC LFCP samples as  $\rm A_{eff} = A_{geo} \times \frac{N_{sel}}{N_{gen}}$, where $\rm A_{geo}$ is the geometric factor of the original particle source used to generate the MC sample, $N_{gen}$ is the total events number of MC sample, and $N_{sel}$ is the number of events passing all of the selections. The final $\rm A_{eff}$ is approximately equal to $\rm 307\,cm^2sr$ for the LFCPs with a mass of 0.511 MeV$/c^{2}$.

\begin{figure*}
	\centering
    \begin{overpic}[width=0.44\textwidth]{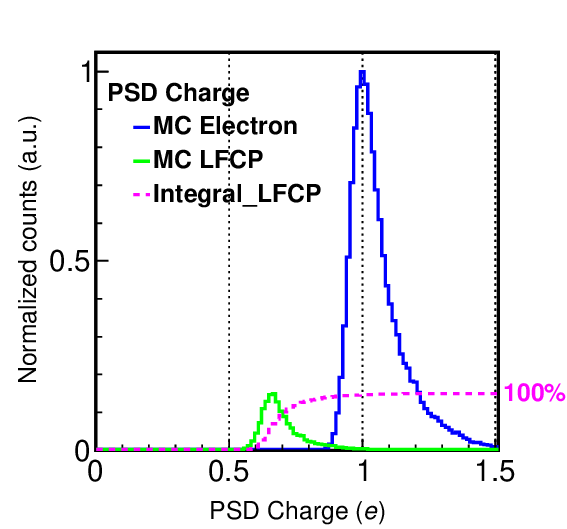}
        \put(2,80){\Large{(a)}}
    \end{overpic}
    \begin{overpic}[width=0.44\textwidth]{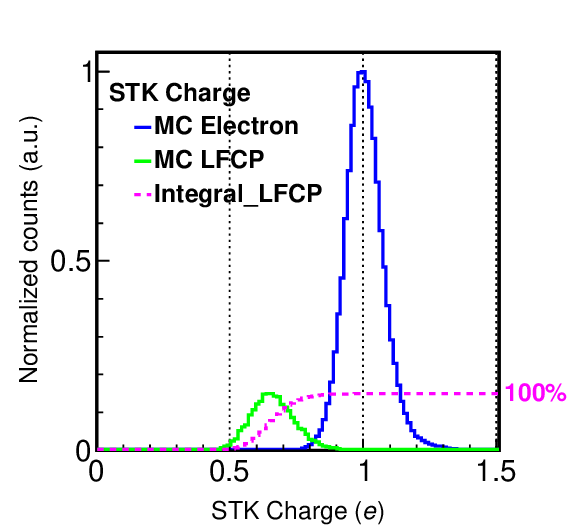}
        \put(2,80){\Large{(b)}}
    \end{overpic}
    \begin{overpic}[width=0.44\textwidth]{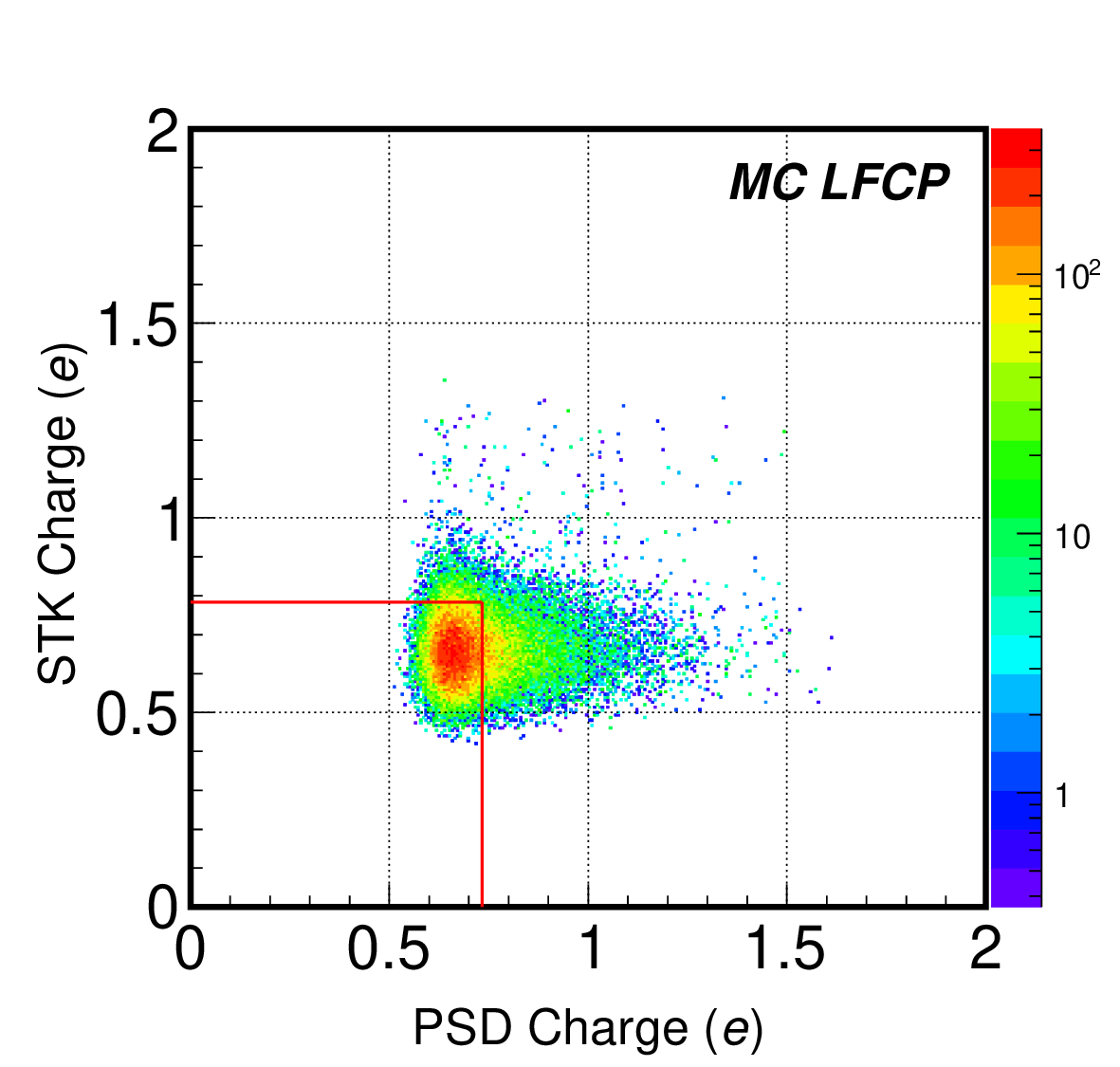}
        \put(2,80){\Large{(c)}}
    \end{overpic}
    \begin{overpic}[width=0.44\textwidth]{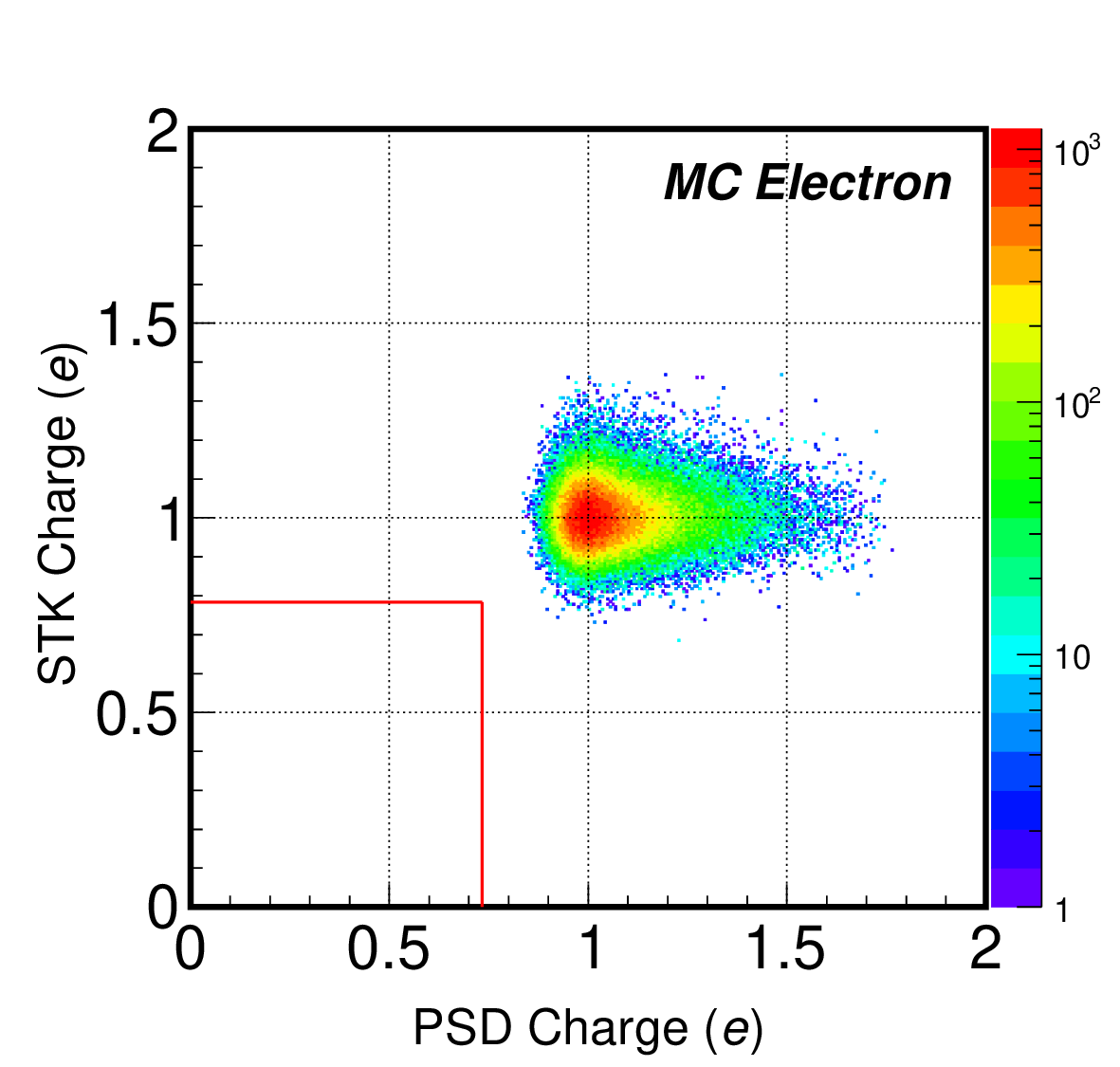}
        \put(2,80){\Large{(d)}}
    \end{overpic}
	  
	\caption{Fig.{2(a)} and Fig.{2(b)} depict the discrimination of MC LFCPs signals and electron backgrounds in the PSD and STK detectors, respectively. The blue lines depict the backgrounds of MC electrons, and the green lines represent the possible distributions of LFCPs signals. The pink integration curves representing the cumulative distributions from the MC LFCPs. The borders were evaluated as  $0.73\,e$ for PSD and $0.78\,e$ for STK, with the efficiencies of 72.4\% and 93.2\%, respectively. 
    The two-dimensional charge distributions of PSD-STK for MC LFCPs and MC electrons are shown in Fig.{2(c)} and Fig.{2(d)}, respectively, with the signal region applied and shown as the red lines. The electrons sample is considered as the main background and away from the signal region. The signal region contains 67.4\% of the MC LFCPs signals.\label{fig:ChrRegion}}
\end{figure*}

$Definition\;of\;the\;signal\;region\;of\;LFCPs$ - 
The PSD and STK charge spectra are fitted using the Landau-Gaussian convoluted function. The resolution $\rm \sigma$ is defined as the square root value of the combined Landau width and Gaussian sigma. The signal region of PSD-STK charge distribution is defined as the most probable value plus 2.5 times $\rm \sigma$. The border values of PSD and STK are $0.73\,e$ and $0.78\,e$, respectively. For the MC LFCPs with 0.511 MeV$/c^{2}$ mass, integrating the charge spectra to the border, the PSD and STK contribute efficiencies of 72.4\% and 93.2\% shown as Fig.~\ref{fig:ChrRegion}(a) and Fig.~\ref{fig:ChrRegion}(b), respectively. The $\rm \epsilon_{region}$, defined as the proportion of events within the signal region, is about 67.4\% for MC LFCPs signals as Fig.~\ref{fig:ChrRegion}(c) shows. The signal region can effectively reject the electron backgrounds as Fig.~\ref{fig:ChrRegion}(d) shows.

\begin{figure}[h]
	\centering
	\includegraphics[width=0.40\textwidth]{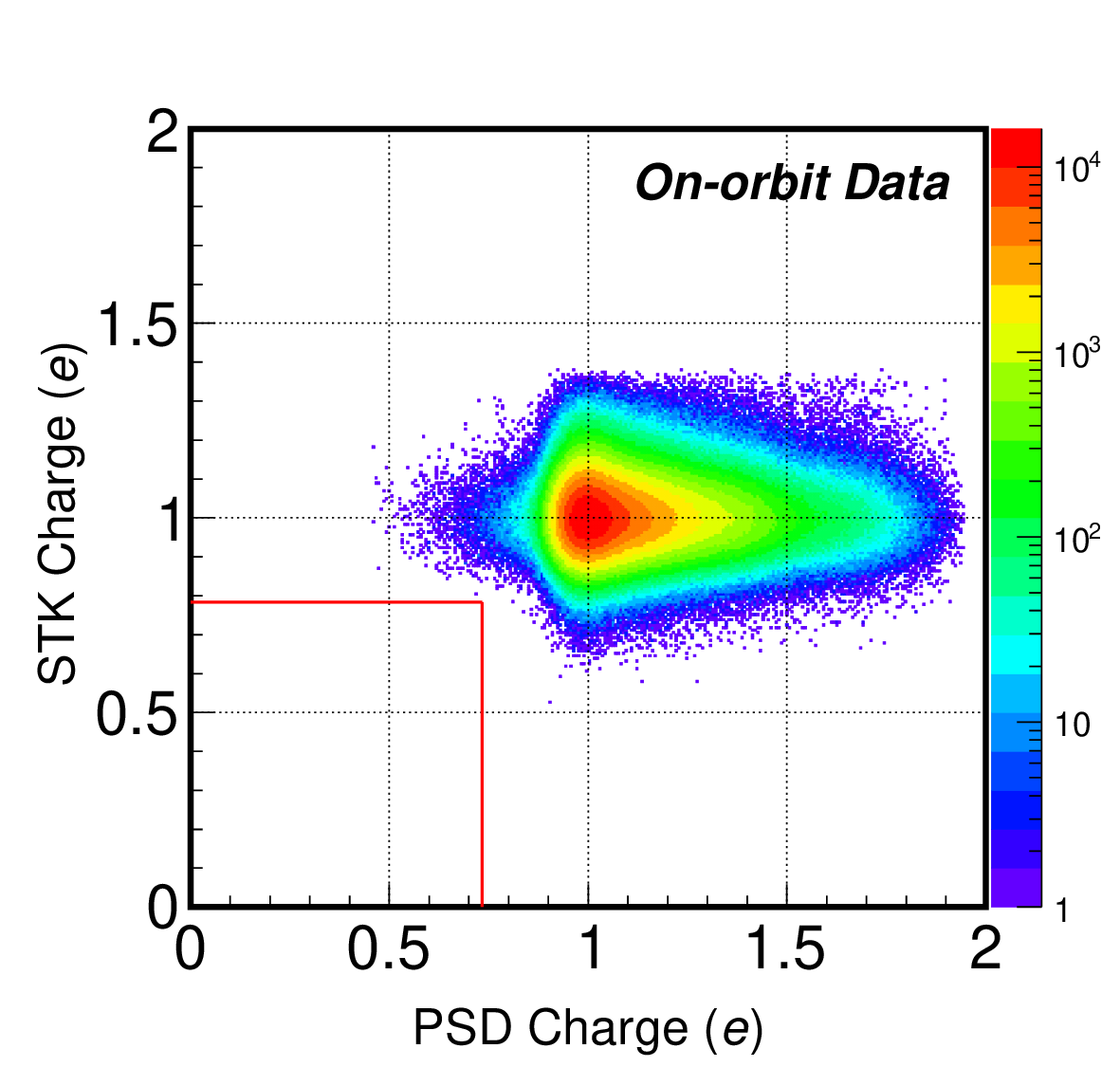}
	\caption{The two dimensional charge distribution of PSD-STK of on-orbit data. The lines define the signal region for LFCP. \label{fig:ChargeSpec_data}}
\end{figure}

The two-dimensional charge distribution of the on-orbit data implementing the signal region is shown in Fig.~\ref{fig:ChargeSpec_data}, which is dominated almost by electron backgrounds, with no LFCP signal observed within the signal region. 

$Systematic\;uncertainties$ - 
The systematic uncertainty originates mainly from the trigger selection, track selection, and charge selection. 
Due to the absence of actual LFCP on-orbit data, half of the difference between the selection efficiencies of MC electrons and the on-orbit data is considered as the systematic uncertainty. 
The combined uncertainty is estimated to be 2.37\%, given by the error propagation Formula~\ref{eq:delta}, where $\delta_{\rm trigger} = 1.08\%$, $\delta_{\rm track} = 1.65\%$ and $\delta_{\rm charge} = 1.31\%$ denote the uncertainties of trigger selection, track selection and charge selection, respectively. 
\begin{equation}
	\rm \delta=\sqrt{\delta_{\rm trigger}^2+\delta_{\rm track}^2+\delta_{\rm charge}^2}.
	\label{eq:delta}
\end{equation}

$Results$- The flux upper limits of $\rm \frac{2}{3}\,e$ LFCPs is given by 
\begin{equation}
	\rm \Phi = \frac{N_{obs}}{T_{exp}A_{eff}\epsilon_{region}(1-\delta)},
	\label{eq:Flux}
\end{equation}

\noindent where $\rm T_{exp}$ denotes the effective exposure time for ten years. $\rm A_{eff}$ is the effective acceptance for LFCPs.
$\rm \epsilon_{region}$ represents the efficiency of signal region for LFCPs. 
And $\rm \delta$ is the total systematic uncertainty. 
$\rm N_{obs}$ is the number of observed LFCPs events from on-orbit data. Since no LFCP event was observed in the signal region as Fig.~\ref{fig:ChargeSpec_data} shows, $\rm N_{obs}$ is taken to be 2.44 \cite{Feldman:1997qc} for the flux upper limit at the 90\% confidence level. 

\begin{figure}[h]
    \centering
	\includegraphics[width=0.5\textwidth]{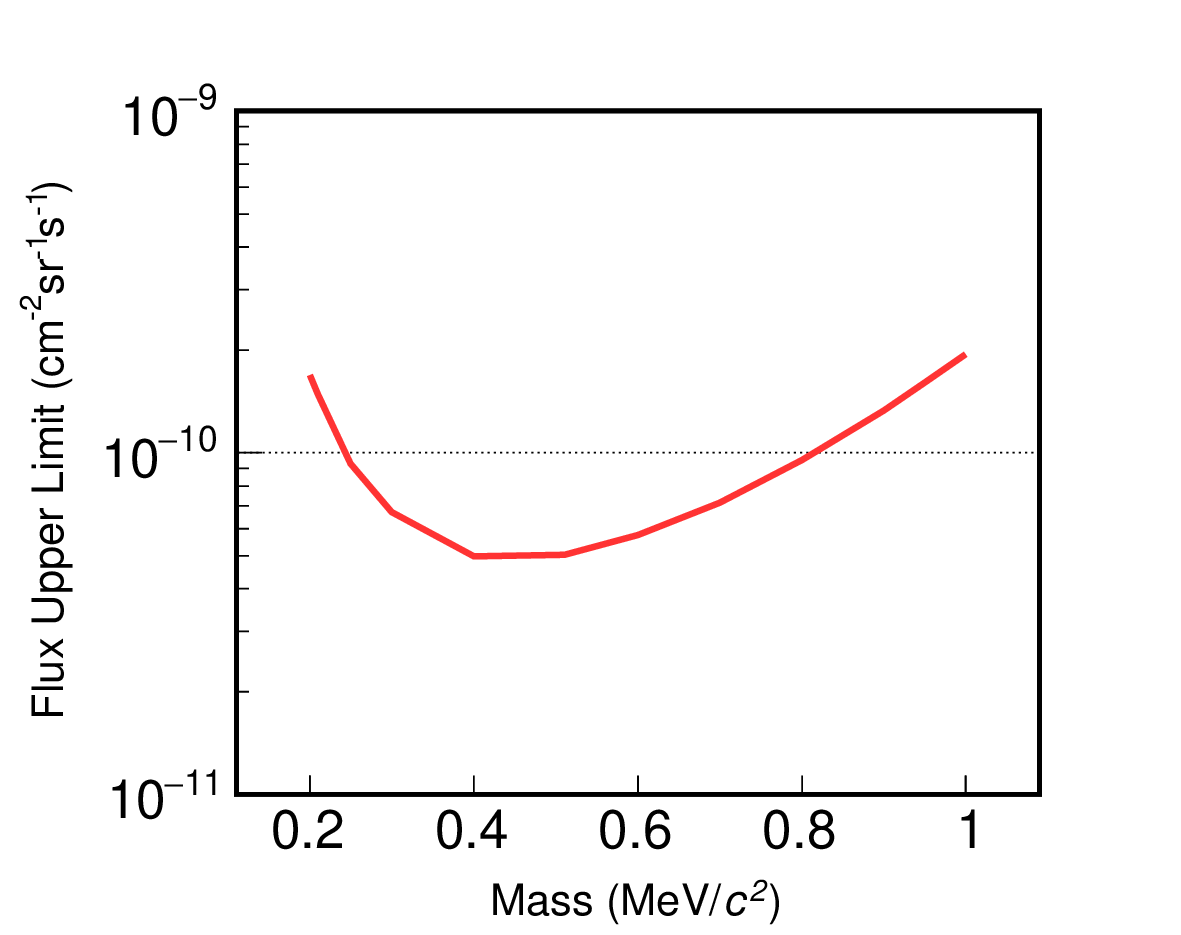}
	\caption{Flux upper limits for LFCP with a mass range from 0.2 to 1 MeV$/c^{2}$.\label{fig:Flux}}
\end{figure}

The flux upper limit for 0.511 MeV$/c^{2}$ LFCPs is determined to be $\rm \Phi_{upper} = 5.0 \times 10^{-11}\,cm^{-2}sr^{-1}s^{-1}$ at the 90\% confidence level with the systematic uncertainties of selections considered. Figure~\ref{fig:Flux} shows the results of the upper limit with particle's mass varying from 0.2 MeV$/c^{2}$ to 1 MeV$/c^{2}$. The LFCPs with a mass around 0.511 MeV$/c^{2}$ have the flux upper limits on the order of $\rm 10^{-11}~cm^{-2}sr^{-1}s^{-1}$.

$Conclusion$ - 
To search for free particles with fractional electric charge, this letter hunts for LFCPs in the sub-MeV mass range using ten-year on-orbit data from DAMPE. The results show that no LFCP have been detected, and the upper limits on the flux of LFCPs are thus derived. Typically, the upper limit at the 90\% confidence level is determined to be $\rm \Phi_{upper} = 5.0 \times 10^{-11}\,cm^{-2}sr^{-1}s^{-1}$ for LFCPs with the mass of 0.511 MeV$/c^{2}$. This is the first search for LFCPs in primary cosmic rays, which supplements the constraint range of similar international experiments.

\begin{acknowledgments}
	
$Acknowledgements$ - 
The DAMPE mission was funded by the strategic priority science and technology projects in space science of Chinese Academy of Sciences (CAS). In China,the data analysis was supported by the National Key Research and Development Program of China (No. 2022YFF0503303) and the National Natural Science Foundation of China (Nos.12220101003, 12588101, 12003076 and 12022503), the CAS Project for Young Scientists in Basic Research (Nos. YSBR-061 and YSBR-092), the Strategic Priority Program on Space Science of Chinese Academy of Sciences (No. E02212A02S), the Youth Innovation Promotion Association of CAS, the Young Elite Scientists Sponsorship Program by CAST (No.YESS20220197), and the Program for Innovative Talents and Entrepreneur in Jiangsu and the NSFC under grant (No.12588101). In Europe, the activities and data analysis are supported by the Swiss National Science Foundation (SNSF), Switzerland, the National Institute for Nuclear Physics (INFN), Italy, and the European Research Council (ERC) under the European Union’s Horizon 2020 research and innovation programme (No. 851103).
 
\end{acknowledgments}

\bibliography{bibfile.bib}

\end{document}


\title{Search for Light-Mass Fractionally Charged Particles with DAMPE - Supplemental Material}
\collaboration{DAMPE Collaboration}
\altaffiliation{dampe@pmo.ac.cn}
\date{\today}
\thispagestyle{empty}

\begin{center}
\vspace*{1cm}

{\LARGE\bfseries Search for Light-Mass Fractionally Charged Particles with DAMPE\\[0.5cm]}

{\Large Supplemental Material\\[1.5cm]}

{\large
\textbf{DAMPE Collaboration}\\[0.3cm]
\url{dampe@pmo.ac.cn}\\[1cm]
}

\today\\[1cm]

\end{center}

\vfill
\titleformat*{\section}{\Large\bfseries\raggedright}
\titleformat*{\subsection}{\large\bfseries\raggedright}

\section{I. LFCPs Simulation}

Except the Ionization effect, the Bremsstrahlung effect also contributes to the energy loss of LFCPs. Both of the ionization and bremsstrahlung effects are proportional to the square of charge value, while bremsstrahlung is also inversely proportional to the square of mass.

\begin{table}[h]
\caption{Processes and models used in LFCPs model. \label{tab:model}}
\begin{tabular}{l|l|l}
	\hline
	Processes & Models & Corrections \\
	\hline
	Ionization & $G4BetheBlochModel$  & Charge  \\
	\hline
	Bremsstrahlung & $G4eBremsstrahlungRelModel$  & Charge, Mass \\
	\hline
	Multiple Scattering & $G4UrbanMscModel$  & Automatic \\
	\hline
\end{tabular}
\end{table}
\subsection{Model establishment}
In DAMPE Software(DMPSW), the Geant4 toolkit~\cite{GEANT4:2002zbu} is used for the particle simulation. Since LFCP is assumed to be an electron-like particle, the model establishment is based on the electron model. The $PhysicsProcess$ including ionization, bremsstrahlung, and multiple scattering is considered.
	
First of all, a particle interface is defined where the mass and charge values are alternative. Then, the models of physics processes are added. These models are succeed from the corespondent ones of electron and corrected with the mass and charge values.

Table~\ref{tab:model} shows the flow chat of the creation of LFCP model. The $G4UrbanMscModel$ for multiple scattering adjusts to the mass and charge of a particle automatically. Thus the corrections mainly focus on the models of ionization and bremsstrahlung. 
	
In the model of $G4BetheBlochModel$, the squared-charge correction is applied to the energy loss $\rm \frac{dE}{dx}$ and the cross section. In the model of $G4eBremsstrahlungRelModel$, the square of charge and inversely square of mass corrections are applied simultaneously to the energy loss and cross section.

\begin{figure}
    \centering
    \begin{overpic}[width=0.45\linewidth]{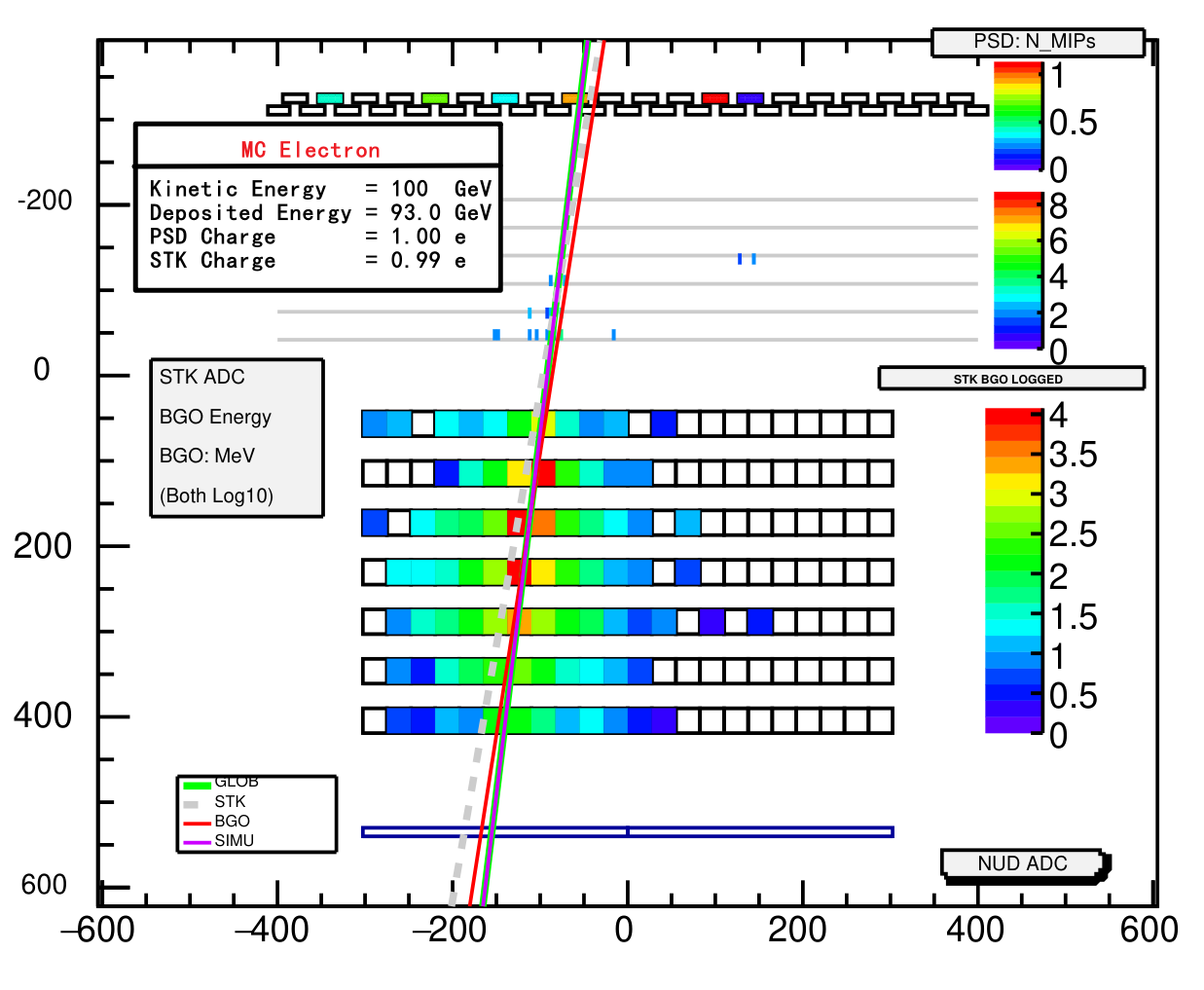}
        \put(-1,75){\Large{(a)}}
    \end{overpic}
    \begin{overpic}[width=0.45\linewidth]{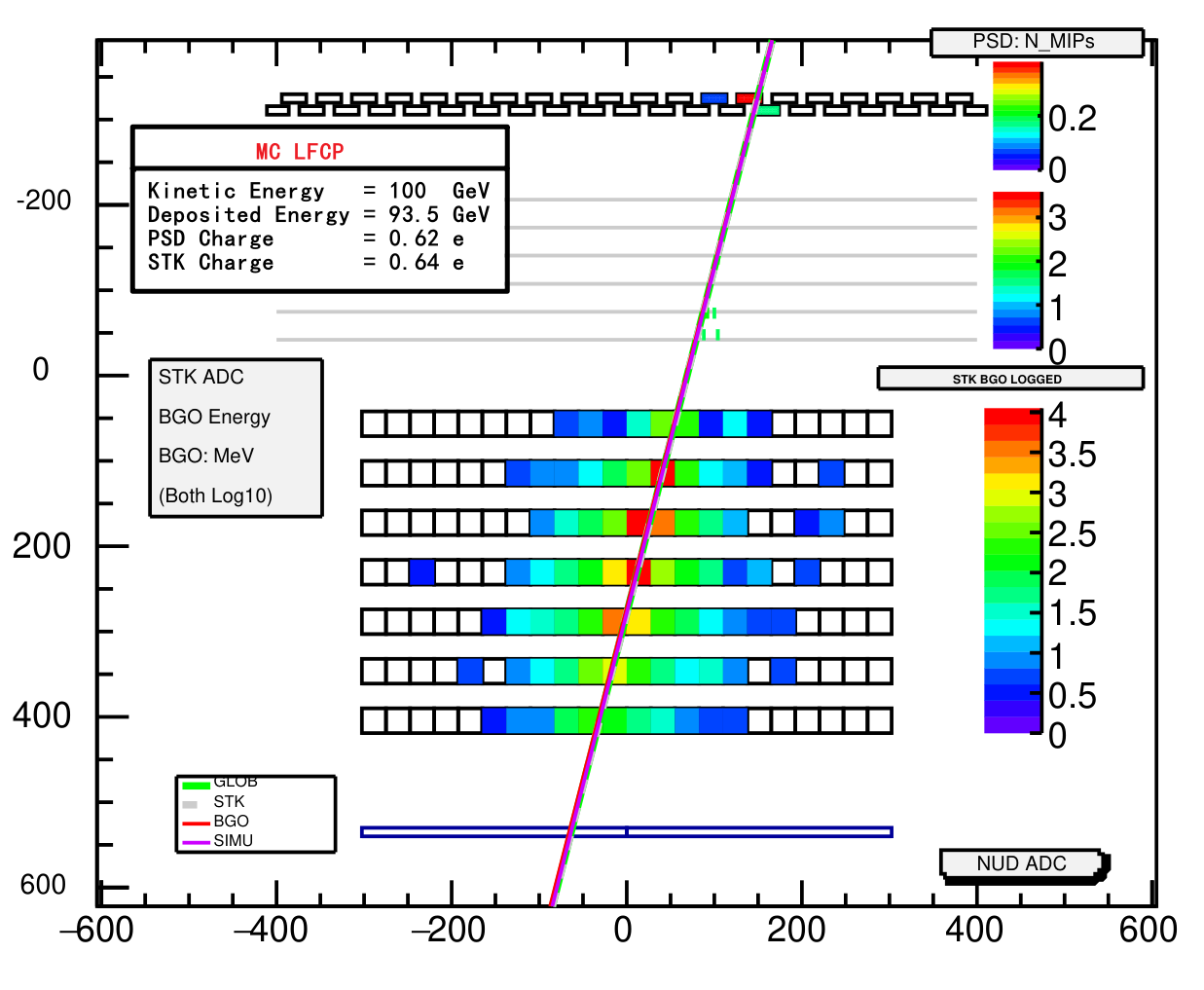}
        \put(-1,75){\Large{(b)}}
    \end{overpic}
    \caption{The responses of DAMPE detector to MC electrons (a) and MC LFCPs (b) at 100 GeV.}
    \label{fig:Simu events display}
\end{figure}
\subsection{Simulation in DMPSW}

In the simulation process, an isotropic hemisphere particle source is adopted. The energy spectrum obeys the power low with index of -1. For each sample, the events number is constrained to be about $10^8$ order. The charge and mass values can be modified in the physics list of LFCPs. 
The responses of DAMPE detector to MC electrons (a) and MC LFCPs with a mass of 0.511 MeV$/c^{2}$ (b) at 100 GeV kinetic Energy are shown in Fig.~\ref{fig:Simu events display}.
About 93\% kinetic energy was deposited in the BGO calorimeter for both MC electron and LFCP events. For LFCP events, the reconstructed charge values of the PSD and STK are $0.62~e$ and $0.64~e$, respectively, while those are $1.00~e$ and $0.99~e$ for electron.

\section{II. High energy trigger}

\begin{figure}[h]

    \centering
	\includegraphics[width=0.75\textwidth]{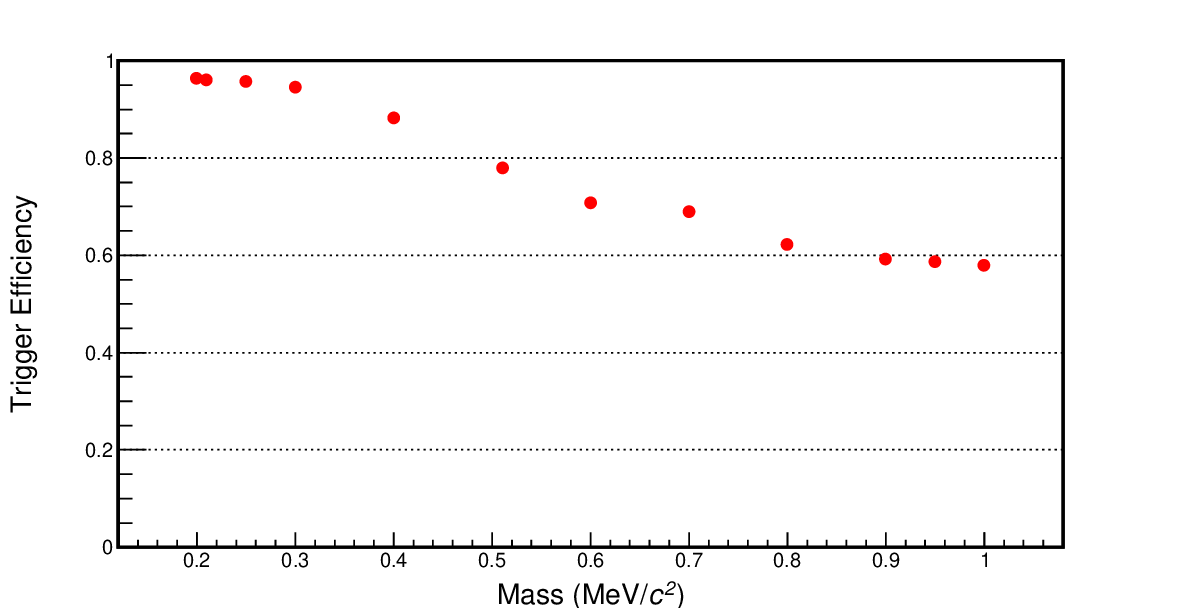}
	\caption{The trigger efficiency of LFCPs samples of different masses.}
	\label{fig:HET_trigger}
\end{figure}

DAMPE has four trigger patterns~\cite{Zhang:2019wkj}, of which the Unbiased Trigger (UNBT) and the High Energy Trigger (HET) are used for this work. 
The efficiency of the trigger, $\rm \epsilon_{trigger}$, is calculated by Formula~\ref{eq:TriggerEff}. 

\begin{equation}
    \rm \epsilon_{HET}=\frac{N_{UNBT\&HET}}{N_{UNBT}}
	\label{eq:TriggerEff}
\end{equation}

\noindent where $\rm N_{UNBT}$, $\rm N_{HET}$, and $\rm N_{UNBT\&\&HET}$ represent the number of events satisfying the UNBT, HET, and both conditions, respectively.
The efficiencies of LFCPs of different masses are shown in Fig.~\ref{fig:HET_trigger}.

\section{III. Charge selection}

The PSD and STK detectors were used to reconstruct the charge of events, a series of criteria are implemented to charge detectors to select the events.

\phantomsection
\subsection{PSD charge selection}
\begin{figure}[h]
    \centering
    \begin{minipage}{0.32\textwidth}
        \centering
        \begin{overpic}[width=\textwidth]{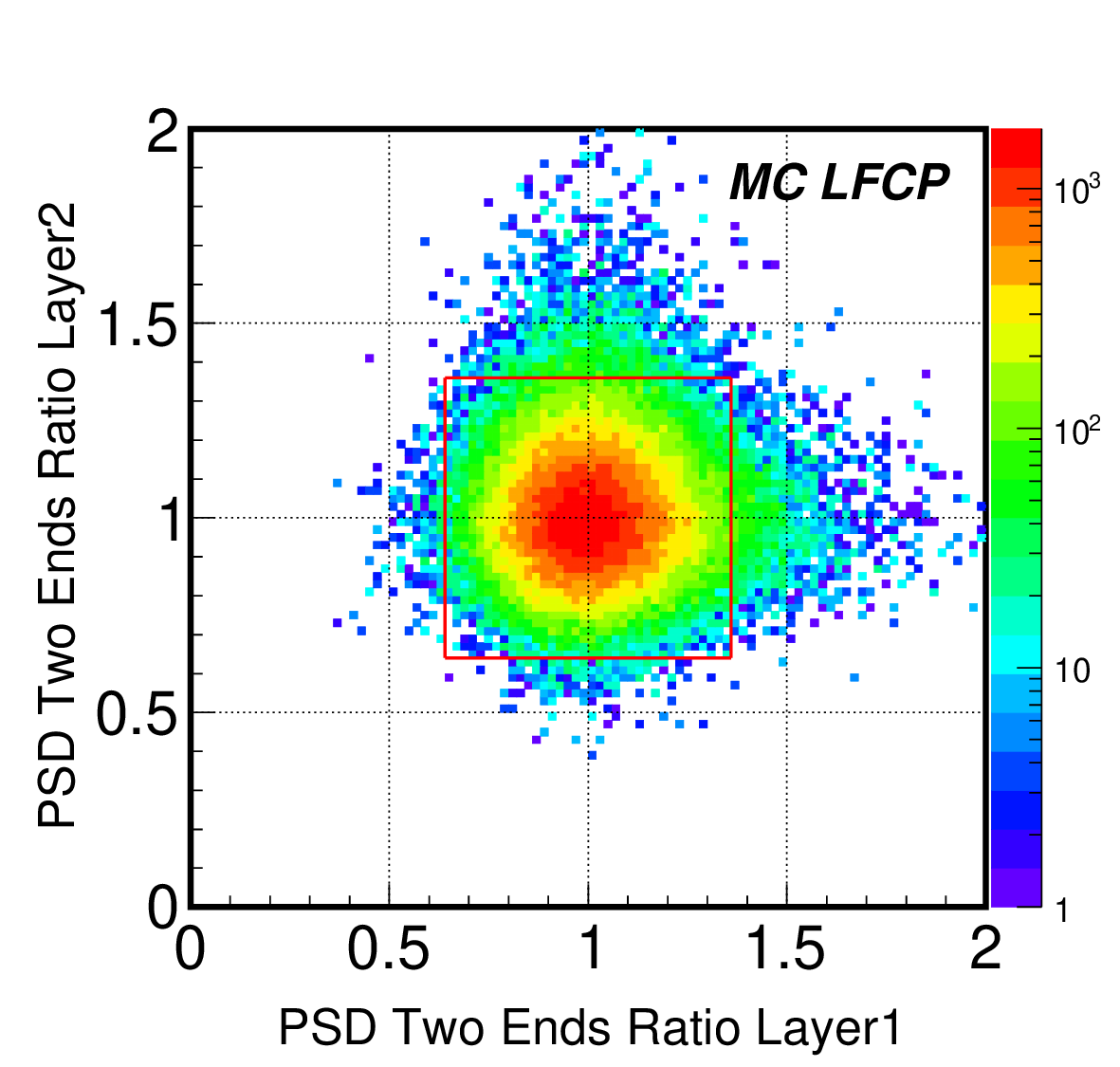}
            \put(0,90){\Large{(a)}}
        \end{overpic}
    \end{minipage}
    \hfill
    \begin{minipage}{0.32\textwidth}
        \centering
        \begin{overpic}[width=\textwidth]{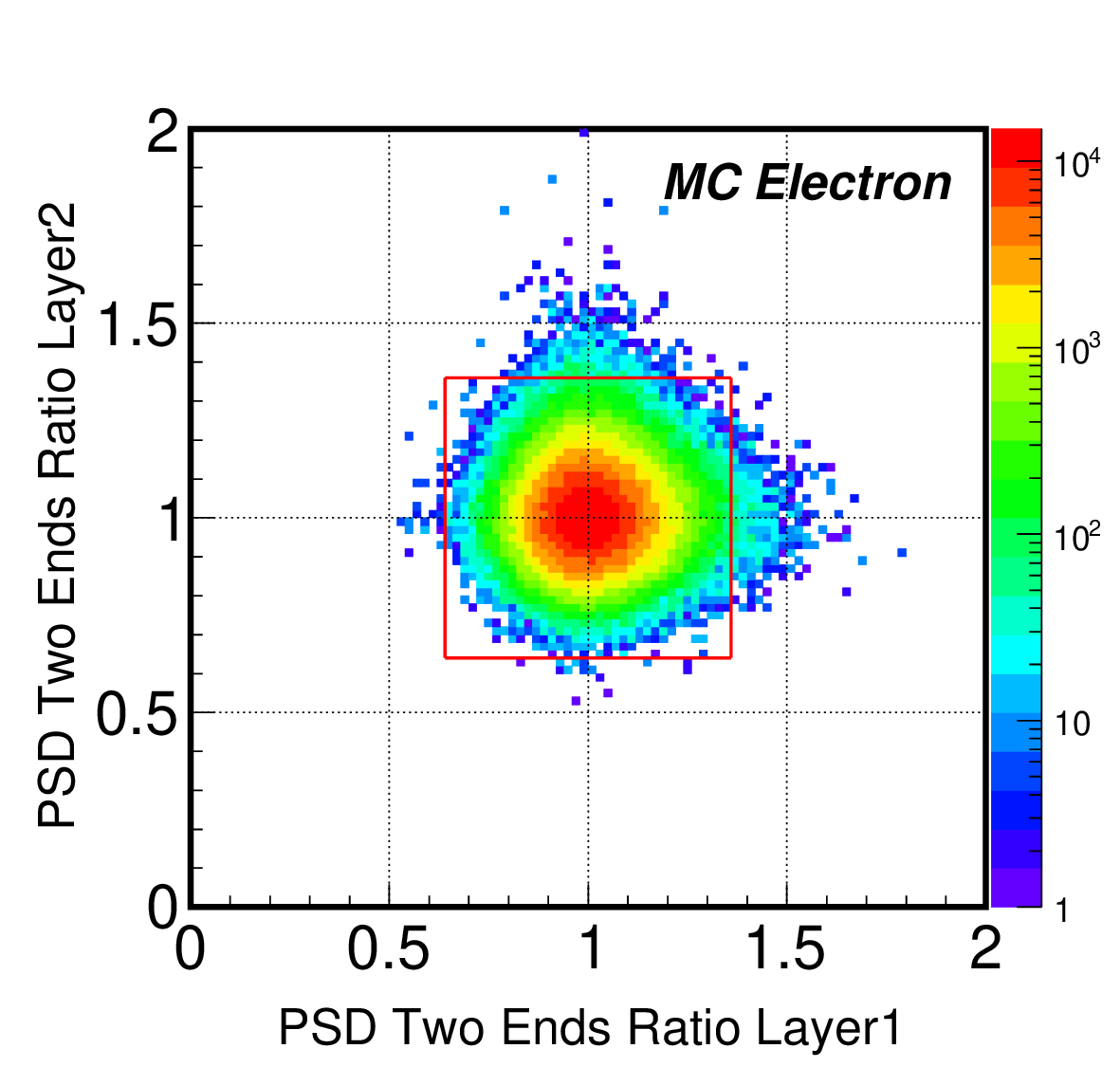}
            \put(0,90){\Large{(b)}}
        \end{overpic}
    \end{minipage}
    \hfill
    \begin{minipage}{0.32\textwidth}
        \centering
        \begin{overpic}[width=\textwidth]{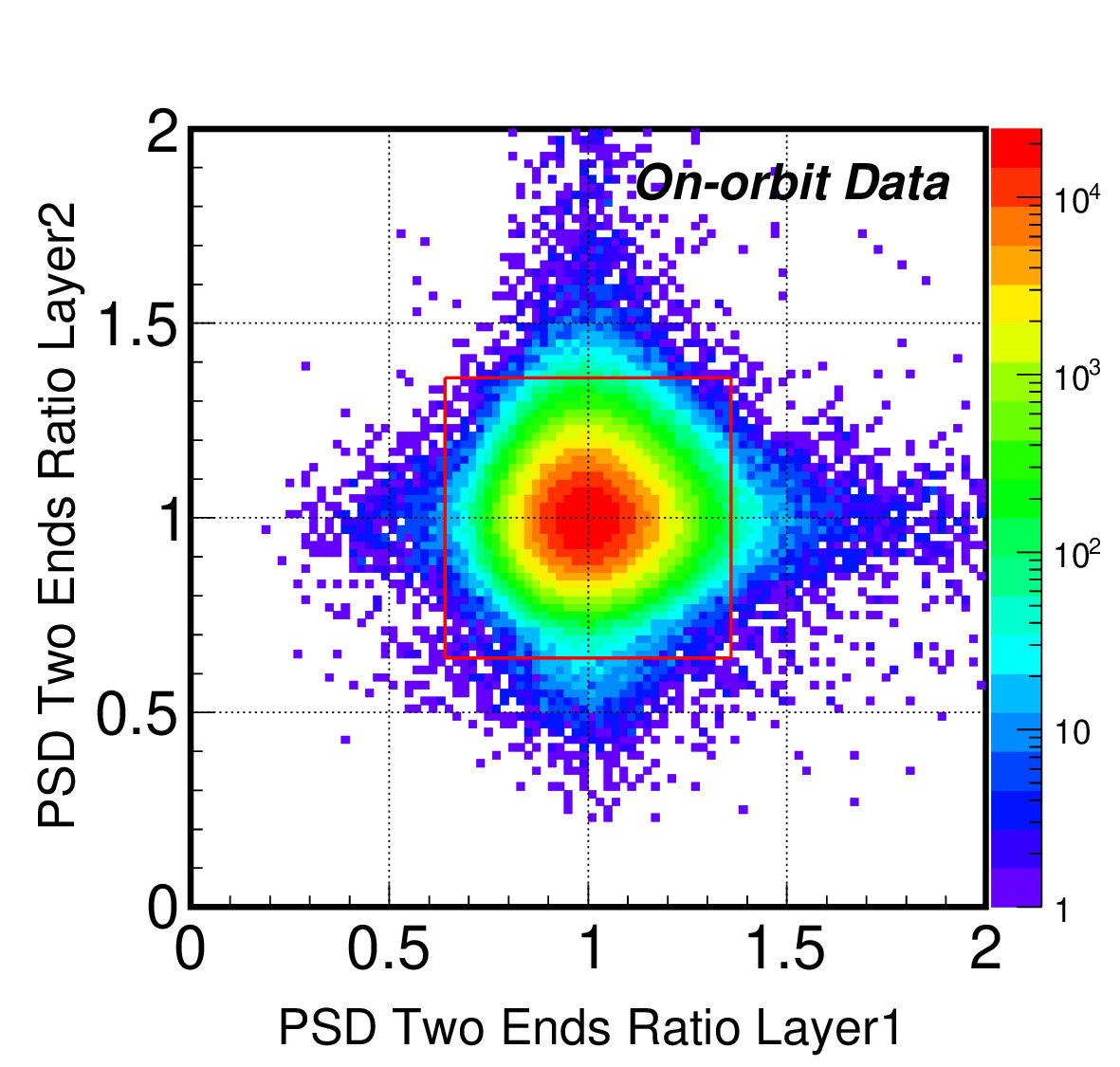}
            \put(0,90){\Large{(c)}}
        \end{overpic}
    \end{minipage}
    \caption{Two-dimensional distributions of two layers of PSD two-end ratio of MC LFCP (a), MC Electron (b) and On-orbit Data (c), where the red square is given by $\rm mean \pm 3\sigma$. \label{fig:PSDRatio}}
\end{figure}

In each layer of PSD detector, the strip with maximum energy deposition around the STK track is selected as the hit strip. The energy deposition is corrected with the path length, light attenuation, and quenching effects.
Furthermore, the tailored conditions are listed as below. 
\begin{itemize}
	\item[(a)] PSD path length: The PSD strips used to reconstruct the charge must be penetrated from top to bottom surfaces.
	\item[(b)] PSD two-end ratio: The signal of one PSD strip is read out by two ends independently. The consistence of two ends' charge is required by constraining the two-end ratio to be inside the range of \textbf{[mean-$\rm 3\sigma$, mean+$\rm 3\sigma$]}, where the $\rm mean$ and $\rm \sigma$ are the parameters of Gaussian fitting.
    The two-dimensional distributions of ends' ratios of two layers of different samples are also shown in Fig.~\ref{fig:PSDRatio} where the red squares denotes the selection ranges.
	\item[(c)] PSD max-bar: The maximum energy deposition among one PSD layer except the one used to reconstruct the charge is required to be less than 10 MeV. 
\end{itemize}

The conditions (a) and (b) are implemented to achieve good resolution, while condition (c) is to exclude the interference of backlash and to improve the accuracy of charge reconstruction.

\newpage
\phantomsection
\subsection{STK charge selection}
\begin{figure}[h]
	\centering
    \vspace{-120pt}
    \includegraphics[width=0.9\textwidth]{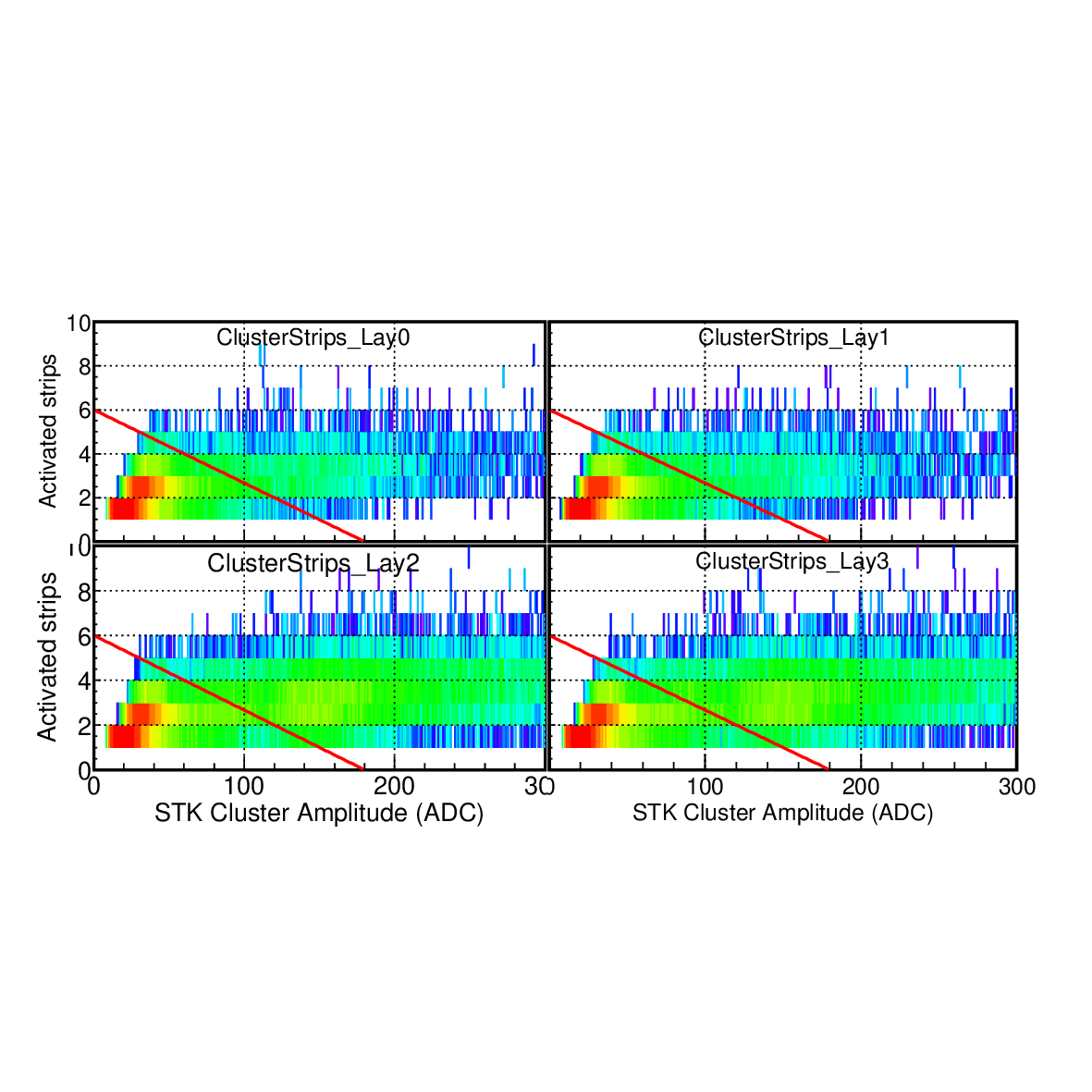}
    \vspace{-100pt}
    \caption{The distributions between the silicon strips and signal amplitudes of first four sub-layers for MC LFCPs. The X-axis is the amplitude of the layer and Y-axis represents the number of activated strips. The red line represents the selecting criteria under which the signals are selected.}
    \label{fig:ClusterStrips}
\end{figure}

There are three 1mm tungsten plates inserted into the STK detector~\cite{Azzarello:2016trx}, which are more likely to induce the pre-shower of electromagnetic particles. Figure~\ref{fig:ClusterStrips} shows the relationships of activated strips and amplitude. With respect to the lower sub-layers, the energy deposition is getting larger.
Thus, the first four sub-layers of the STK are used to reconstruct the charge.

The silicon strips are sensitive to the energy deposition of secondary particles, especially to the LFCPs with charge less than a $e$. To purify the signals and eliminate the influence of secondary particles, some criteria are implemented. 

\begin{itemize}
    \item[(a)] Only signals from the first four detector sub-layers are selected for charge reconstruction. Furthermore, to ensure the full deposition of energy, the events that hit the edges of silicon sensors are excluded.
	\item[(b)] A selection is applied based on the relationship of signal's amplitude and the number of activated silicon strips, which must satisfy Formula~\ref{eq:NStrips}. 
	\begin{equation}
		\rm ClusterAmplitude \times \frac{6}{180} + NStrips < 6.
		\label{eq:NStrips}
	\end{equation}
	\item[(c)] At least three signals should be remained after the last two steps. The effective signals must satisfy good consistency, requiring their variance to be less than 250. 
\end{itemize}

\begin{figure}[h]
    \centering
	\includegraphics[width=0.75\textwidth]{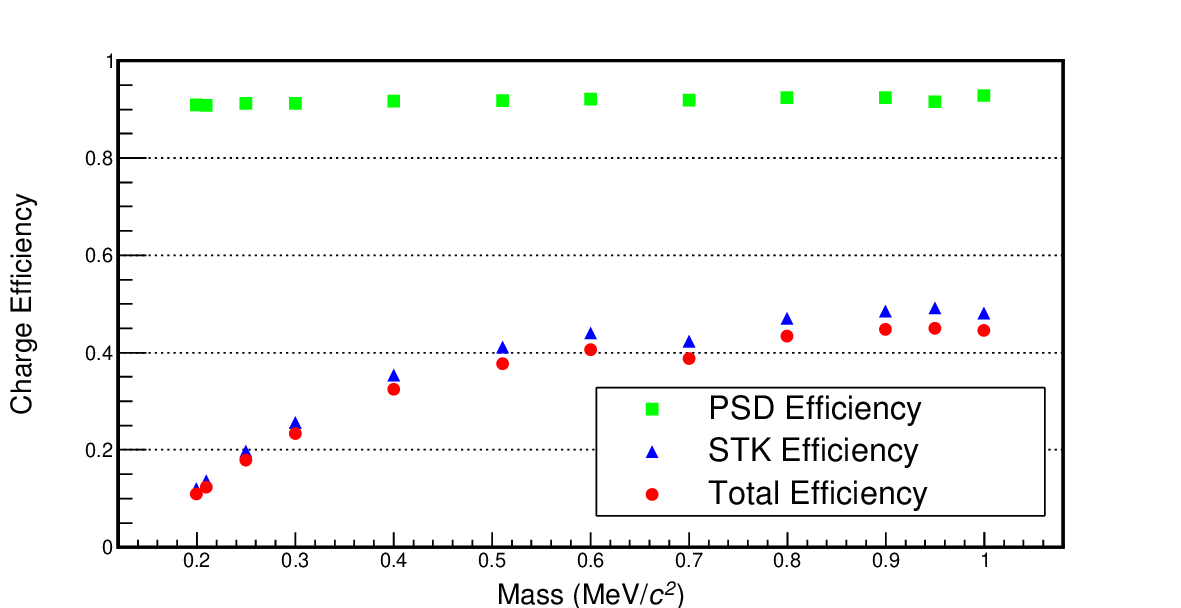}
	\caption{The efficiencies of the PSD (green square), STK (blue triangle) detectors and the total efficiency (red dot) of LFCPs samples of different masses.}
	\label{fig:charge_eff}
\end{figure}

Finally, the average values of two-layer PSD detector and effective signals of STK sub-layers are used as the final charges of $\rm Q_{PSD}$ and $\rm Q_{STK}$ charges, calculated as Formula~\ref{eq:CalCharge}. Figure~\ref{fig:charge_eff} shows the efficiency of charge reconstruction. Since the selections of STK strictly constrain the pre-shower events, the efficiencies are relatively lower than those of PSD. The total efficiencies of charge reconstruction are also shown in FIG.~\ref{fig:charge_eff}

\begin{subequations}
	\begin{equation}
		\rm Q_{PSD} = \frac{Q_0+Q_1}{2},
		\label{eq:PSDCharge}
	\end{equation}
	\begin{equation}
		\rm Q_{STK} = \dfrac{\Sigma^{N}_{i=1}Q_i}{N}.
		\label{eq:STKCharge}
	\end{equation}
    \label{eq:CalCharge}
\end{subequations}

\newpage

\section{IV. Systematic uncertainties}

The effective acceptance results from the combined effect of all the selection criteria. 
The systematic uncertainties originate from the contributions of trigger selection, track selection, and charge selection. 
Due to the absence of actual LFCP orbit data, half of the difference between the MC electrons and the on-orbit data is considered as the systematic uncertainty associated with a given selection.

When calculating the efficiency of certain selection, all of the other selections are used to select a pure sample,
thereby reducing the influence of other selections and providing a more precise estimate.
The uncertainty is performed with Formula~\ref{eq:sysUncertainty}.

\begin{equation}
	\rm \delta_{sel} = \frac{Eff_{Data}-Eff_{MCElectron}}{2 \times Eff_{MCElectron}}.
	\label{eq:sysUncertainty}
\end{equation}

Table \ref{tab:EffAndSys} illustrates the systematic uncertainties and the efficiencies of correspondent selection.

\begin{table}[h]
\caption{Systematics and correspondent selection efficiencies. \label{tab:EffAndSys}}
\begin{tabular}{l|l|l|l}
	\hline
	Efficiencies & MC Electrons & On-orbit Data & Systematics ($\rm \delta_{sel}$) \\
	\hline
	Trigger & 92.16 \% & 90.17 \% & 1.07 \% \\
	\hline
	Track & 97.84 \% & 94.60 \% & 1.65 \% \\
	\hline
	Charge & 36.56 \% & 35.50 \% & 1.31 \% \\
	\hline
\end{tabular}
\end{table}

As the mass increases, the physical behavior of LFCPs in the detector becomes increasingly similar to a Minimum Ionizing Particle. Consequently, the algorithm presented in this paper is not suitable for LFCPs with an excessively large mass. Meanwhile, for very small mass, due to the increasing influence of bremsstrahlung effect, LFCPs exhibit poor charge resolution in the STK. 

\begin{figure}[htbp]
	\centering

    \begin{overpic}[width = 0.75\textwidth]{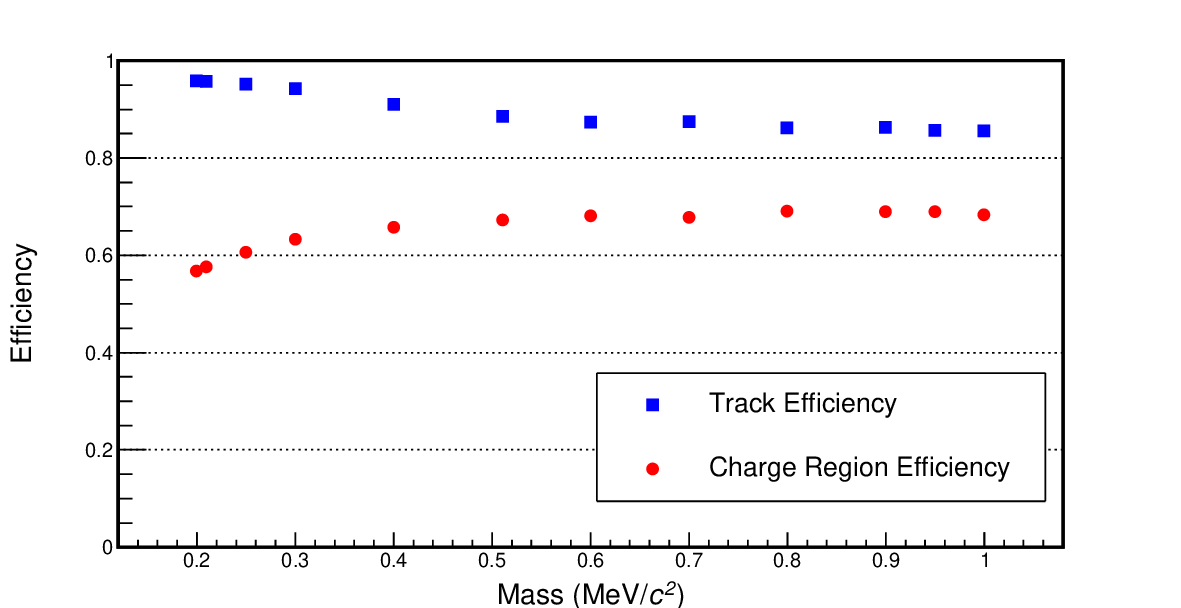}
        \put(1,45){\Large{(a)}}
    \end{overpic}
    \begin{overpic}[width = 0.75\textwidth]{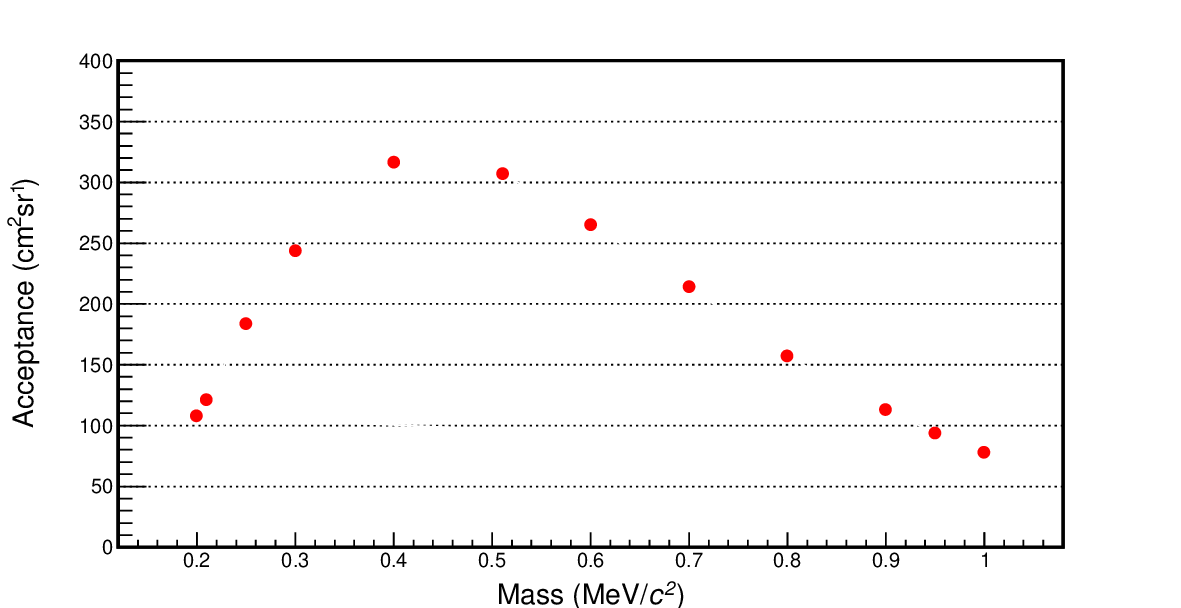}
        \put(1,45){\Large{(b)}}
    \end{overpic}
	\caption{(a) The efficiencies of track selection (blue square) and signal region (red dot), and (b) the final effective acceptance of LFCPs of different masses.}
	\label{fig:MassScanning}
\end{figure}

The efficiencies of track selection and signal region are shown in Fig.\ref{fig:MassScanning}(a), as well as the effective acceptance shown in Fig.\ref{fig:MassScanning}(b).

\newpage
\bibliography{bibfile.bib} 